\documentclass[5p,times,twocolumn]{elsarticle}
\usepackage{macros}
\begin{document}
\begin{frontmatter}
\title{Education in the Era of Neurosymbolic AI}
\author[wsu]{Chris Davis Jaldi}
\ead{jaldi.2@wright.edu}
\author[l3s]{Eleni Ilkou}
\ead{ilkou@l3s.de}
\author[ufl]{Noah Schroeder}
\ead{schroedern@ufl.edu}
\author[wsu]{Cogan Shimizu}
\ead{cogan.shimizu@wright.edu}
\address[wsu]{Wright State University, US}
\address[l3s]{L3S Research Center, Leibniz University Hannover, DE}
\address[ufl]{University of Florida, US}
\begin{abstract}
Education is poised for a transformative shift with the advent of neurosymbolic artificial intelligence (NAI), which will redefine how we support deeply adaptive and personalized learning experiences. NAI-powered education systems will be capable of interpreting complex human concepts and contexts while employing advanced problem-solving strategies, all grounded in established pedagogical frameworks. This will enable a level of personalization in learning systems that to date has been largely unattainable at scale, providing finely tailored curricula that adapt to an individual's learning pace and accessibility needs, including the diagnosis of student understanding of subjects at a fine-grained level, identifying gaps in foundational knowledge, and adjusting instruction accordingly. In this paper, we propose a system that leverages the unique affordances of pedagogical agents—embodied characters designed to enhance learning—as critical components of a hybrid NAI architecture. To do so, these agents can thus simulate nuanced discussions, debates, and problem-solving exercises that push learners beyond rote memorization toward deep comprehension. We discuss the rationale for our system design and the preliminary findings of our work. We conclude that education in the era of NAI will make learning more accessible, equitable, and aligned with real-world skills. This is an era that will explore a new depth of understanding in educational tools.
\end{abstract}
\end{frontmatter}
\section{Overview \& Motivation}
\label{sec:intro}
\begin{figure*}[t]
    \centering
    \includegraphics[width=0.75\linewidth]{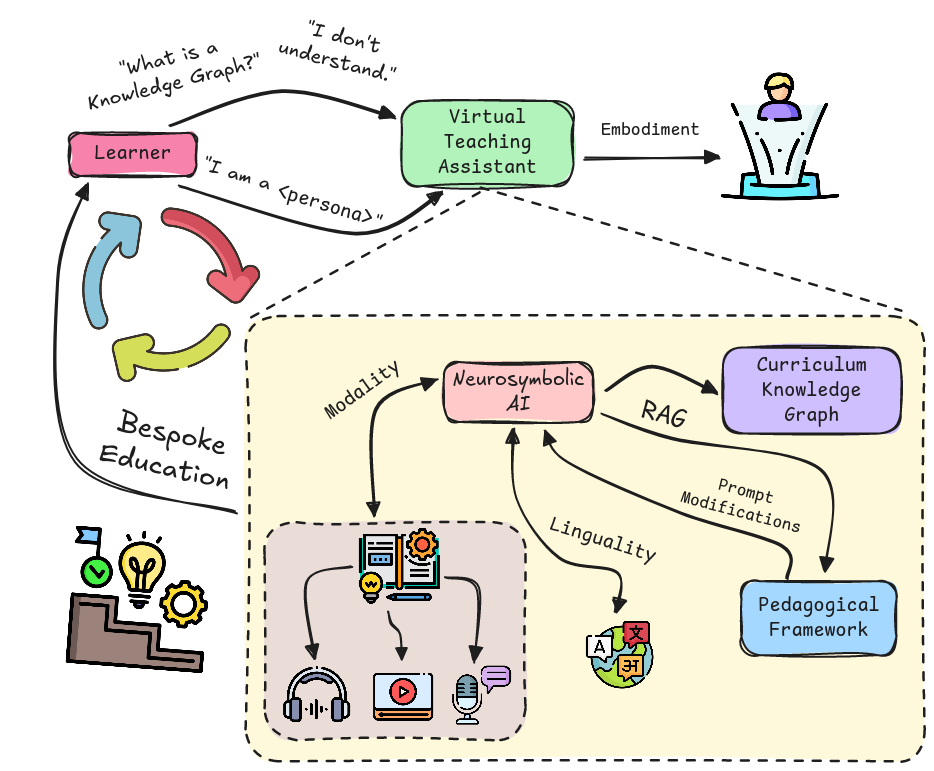}
    \caption{An overview of the iterative learning-intervention cycle enabled by the integration of Neurosymbolic AI in education.}
    \label{fig:napa-ov1}
\end{figure*}
The integration of technology into education -- especially artificial intelligence (AI) -- has accelerated rapidly, transforming traditional learning environments. Yet, many existing educational technologies face significant limitations and still fall short in delivering deeply personalized and adaptive learning experiences \cite{gligorea2023adaptive}. While tools such as intelligent tutoring systems (ITS), learning management systems (LMS), and (adaptive) educational apps (e.g., \cite{harley2015multi, nesbit2014effective}) have improved access to knowledge and enhanced engagement through better accessibility and resource availability, they often lack the depth of understanding and contextual adaptability required for deeply personalized instruction~\cite{prather2023robots}. Traditional AI models, particularly those without advanced reasoning capabilities, have a limited scope when it comes to recognizing and addressing the unique learning pathways of individual learners, which poses challenges for tailoring educational content to specific learner needs and providing real-time, context-aware feedback. This gap in adaptability and personalization presents a compelling opportunity that can be addressed by an emerging field in AI: Neurosymbolic AI.

Neurosymbolic AI (NAI) is a hybrid approach that combines the strengths of symbolic AI and neural systems, enabling a better big-picture understanding of complex human concepts and contextual subtleties \cite{hitzler2022neuro,garcez2023neurosymbolic}. By integrating pattern recognition abilities driven by neural systems with the logical reasoning strengths of symbolic AI, NAI offers a comprehensive approach to human learning. In the context of education, this synergy opens up exciting possibilities for the development of pedagogical agents (PAs) -- virtual on-screen characters designed to support and enhance the learning process~\cite{siegle2023twenty, zhang2024pedagogical}. The incorporation of NAI, along with NAI-augmented pedagogical agents (NaPAs), holds promise for transformative advancements in personalized learning systems. These agents are envisioned to support personalized learning through intelligent context interpretation, dynamic interaction, and personalized curriculum development. NaPAs could surpass the limitations of current AI tools that are broadly implemented in educational settings, by  introducing a hybrid human-AI interface that bridges the NAI with the human intuition. The significance of integrating NAI with educational technology becomes especially clear when examining one of the foundational components involved -- large language models (LLMs).

LLMs (e.g., OpenAI's ChatGPT \cite{site:gpt-4o} or Meta's Llama \cite{site:llama}) have demonstrated impressive capabilities in understanding and generating human-like text, facilitating interactions that feel natural and conversational \cite{brown2020language, huynh2023understanding}. Despite these advancements, LLMs struggle with reasoning tasks and often lack intrinsic comprehension of logical relationships, causality, or deep contextual understanding, even during optimized retrieval through various prompting techniques \cite{chen2023unleashing, white2023prompt}. However, in an NAI context, we can leverage other strategies to address these gaps. Specifically, knowledge graphs (KGs) provide a structured approach to representing knowledge through interconnected nodes and edges, where each node represents a concept or entity, and edges capture their relationships~\cite{hitzler-cacm,hogan2021knowledge,kgs}. By organizing domain-specific knowledge, KGs help AI models recognize interconnections between concepts, leading to a more comprehensive and contextually accurate representation of information. This structure is especially crucial in educational settings, where understanding the context and relationships between ideas is essential for building a learner’s knowledge base~\cite{chen2018knowedu}. Educational KGs, when integrated with student models~\cite{student-model} and personal KGs \cite{pkgs}, facilitate customized learning based on both the learner’s prior knowledge and identified knowledge gaps.

One of the major limitations of many existing educational technologies is their reliance on generic, one-size-fits-all curricula \cite{murray2004issues}. The adaptive capabilities of NaPAs will fundamentally transform this landscape with their advanced ability to understand learner profiles, delivering educational content in a manner that is deeply personalized to each learner’s pace, and skill level, as well as current and future needs.

For example, consider two human learners: one, a computer science student already well-versed in knowledge engineering, and the other, a stakeholder with a background in problem-solving and business. A typical tutoring system may provide the same fixed information of the whole solution to both learners, but lacks the capacity to address the distinct conceptual challenges each might face~\cite{vanlehn2011relative}. NaPAs, on the other hand, are identifying patterns of learner behavior that suggest a deeper misunderstanding -- such as confusion around concept ambiguity -- corresponding to the learners' background knowledge, providing more personalized instruction to reinforce specific concepts.

Finally, we note that NaPAs are poised to have global social impact.
In alignment with the United Nations Sustainable Development Goal~4~\cite{site:un-sdg4}, which aims to ensure inclusive and equitable quality education for all, NaPAs will enable personalized education and learning, and powered by deployable cutting-edge technology. This will be accessible in human-resource-strapped environments and underprivileged populations, and offer a variety of adaptive content to fit the stigma against special education and learning difficulties.

In this paper, we discuss such a hybrid system that combines symbolic knowledge from educational KG and pedagogical frameworks with neural systems and LLMs for deeper personalization and accessibility. Figure!\ref{fig:napa-ov1} shows an envisioning of the NaPA. The system recognizes nuances in learner performance and identifies specific gaps in learner knowledge, as well as identifies the areas they know well. It has the capacity to generate conversational instruction in various formats using evidence-based pedagogies, understand multimodal inputs, translate information between modalities, and present information as multimodal outputs, resulting in deeply personalized instruction for each learner. This NAI system and capabilities would act as the ``back-end'' for a personalized, embodied pedagogical agent as the ``front-end''. 

In Section~\ref{sec:foundations}, we offer insights into the foundational concepts underlying the system and present the case for integrating NAI into educational settings.
Next, Section~\ref{sec:nai-pa} provides an overview of the challenges, limitations, and future directions for NAI in education, addressing the obstacles in implementing NAI-powered educational tools and emphasizing the ethical and practical considerations essential to this rapidly evolving field.
We briefly report key (preliminary) findings and their implications for practice in Section~\ref{sec:results}. 
Finally, in Section~\ref{sec:conc} we conclude. 

\section{Foundations}
\label{sec:foundations}
To better understand the potential of NAI in education, it is essential to explore the foundational components of NAI in educational contexts: KGs, LLMs, and PAs.

\subsection{Large Language Models in Education}
\label{ssec:back-ellm}
LLMs are advanced AI systems trained on vast datasets with remarkable capabilities in natural language processing tasks, including text completion, translation, and summarization. In educational contexts, LLMs have been used to develop ITS, automate grading, and generate educational content~\cite{huang2024generating}. 

However, LLMs face limitations, particularly in reasoning and contextual understanding. Their reliance on data recognition without explicit knowledge representation can lead to inaccuracies, especially in complex problem-solving scenarios \cite{li2024long}. 

Although retraining LLMs -- an often computationally expensive process -- and fine-tuning them \cite{jeong2024fine} are viable strategies, LLMs are ``black box'' methods that are generally not readily explainable or interpretable by the layperson and thus, a significant gap remains \cite{llms-explainability}. In short, LLMs are not based upon a ``ground truth'' of knowledge representation and structure, but instead an approximation of knowledge through language, which can result in ``hallucinations'' where they create factually false information. This is particularly dangerous in educational contexts as at the present time, many LLMs have a tendency to present information as though it is fact, even if it is a hallucination \cite{gpt-personality}. This limitation underscores the need to incorporate structured knowledge representations, such as KGs, to enhance LLM effectiveness in educational applications.

\subsection{Knowledge Graphs in Education}
\label{ssec:back-ekg}
KGs are structured representations of information, where entities (nodes) are connected by relationships (edges), forming a network of knowledge. In education, KGs have been employed to organize and represent domain-specific knowledge, facilitating personalized learning pathways and intelligent content recommendations. For instance, educational KGs can map out the relationships between concepts in a curriculum, enabling adaptive learning systems to identify prerequisite knowledge and tailor instruction to individual learners' needs~\cite{ain2023automatic,chen2018knowedu,qu2024survey}. 

Meanwhile, personal KGs are structured resources containing information about entities and activities related to a user \cite{balog2019personal,ilkou2022personal}. In this context, personal KGs can capture, for a particular learner, their learning analytics, personalized recommendations, and question-answering to improve the learner experience and engagement~\cite{ain2024learner, DBLP:journals/tlt/IlkouTFT23, liu2024question}. By providing a structured framework, KGs enhance the interpretability and reasoning capabilities of AI systems, addressing some of the limitations inherent in LLMs. 

\subsection{Pedagogical Agents}
\label{ssec:back-pa}
While LLMs and KGs create an AI foundation for an effective learning system, the system can be enhanced by evidence-based pedagogical features. One such affordance that complements the conversational presentation style of LLMs are PAs. PAs are embodied software entities designed to facilitate learning by interacting with learners in educational environments \cite{siegle2023twenty}. They can vary in appearance from human-like characters to non-human-like characters (for example, a panda \cite{wu2023zoomorphic}, parrot~\cite{atkinson2002optimizing}, or other non-human character). Having been researched for more than twenty years \cite{siegle2023twenty}, PAs are not a new phenomenon, and research has widely shown that students learn more with PAs than in learning systems without PAs \cite{castro2021effectiveness,schroeder2013effective, wang2023effects}, and that PAs also can improve motivation \cite{wang2023effects}. There is also a plethora of research examining how specific design features of PAs subsequently influence how effective they are at improving learning and how they are perceived by the user \cite{heidig2011pedagogical, zhang2024pedagogical}. For example, PAs can take various roles in a learning environment, such as acting as an information source, coaching, or demonstrating or modeling tasks \cite{clarebout2002animated, schroeder2015persisting}, and are frequently implemented to enhance engagement, provide feedback, and support personalized learning. It is not surprising then that the use of PAs in learning systems continues to increase.

\subsection{Pedagogical Frameworks}
\label{ssec:back-pf}
There are two primary reasons why PAs are thought to be particularly effective learning tools: the social conversation schema they activate in the learner, and the pedagogy they embody. 

It is well known that humans treat computers as though they are conversational partners \cite{reeves1996media}. This has been integrated into theories of how we learn with and from technology for more than 20 years~\cite{mayer2003social}, and recently a more comprehensive theory has been proposed. This recent theory, called the Cognitive-Affective-Social Theory of Learning in Digital Environments (CASTLE), proposes that social aspects of a learning environment (such as a PA) activate social schemata in the learner, which then influences their learning-related processes (e.g., meta-cognitive and motivational processes), thus influencing learning \cite{schneider2022cognitive}. In short, theory and empirical evidence show us that integrating a PA into a learning environment improves learning and motivation, likely due to the initiation of social processes within the learner from engaging with the PA as a conversational partner.

As noted, initiating the social conversation schema is only part of the reason why PAs are effective learning tools. Another reason is the pedagogy they embody. To maximize their positive effects on learning, PAs must be designed in ways that are known to be effective for teaching and learning. While there are a huge number of possible strategies PAs can use, we provide one such example: retrieval practice. 

Retrieval practice is the active and intentional recall of information from memory \cite{karpicke2011retrieval,roediger2011critical}. One form of retrieval practice is known as the testing effect, where students take a practice test \cite{adesope2017rethinking}. The testing effect has been found to be quite effective for improving learning, with a meta-analysis of over 15,000 participants' scores across 272 studies finding an effect size of \textit{g} = .70, \textit{p} < .05. This is consistent with a review of retrieval practice more broadly that found consistent positive effects on learning~\cite{agarwal2021retrieval}. Retrieval practice can be leveraged alongside strategies such as spaced repetition \cite{carpenter2022science}. This technique involves reviewing information at various intervals to enhance long-term retention. This reinforces memory consolidation \cite{kang2016spaced}, and a recent meta-analysis showed the strength of this technique for learning \cite{latimier2021meta}. Furthermore, one can leverage interleaving during spaced retrieval practice. This strategy involves mixing different topics or types of problems within a single study session rather than focusing on one topic at a time. Interleaving has been shown to improve problem-solving skills and adaptability by promoting discrimination between different concepts \cite{kang2016spaced}, although meta-analytic evidence suggests that the effectiveness of interleaving may depend on the learning materials and task \cite{brunmair2019similarity, firth2021systematic}. Notably, retrieval practice and related techniques can be accomplished without the use of a PA. However, the PA adds a social element to the learning environment which can influence learning (see CASTLE theory), and these strategies provide just a few examples of types of evidence-based pedagogy a PA can use.

\section{The Case for Neurosymbolic AI powering Pedagogical Agents}
\label{sec:nai-pa}
Neurosymbolic AI offers comprehensive approaches for advancing educational technology, specifically by integrating the complementary strengths of neural systems with the rigorous, structured reasoning of symbolic AI. This hybrid model is particularly advantageous in educational settings, where it enables PAs to better interpret and respond to complex, nuanced information while adapting instruction to the unique needs of each student. Moreover, these hybrid systems have the ability to work in multiple modalities -- such as text, audio, and video -- which opens new possibilities to create a more immersive and flexible environment for an enhanced learning experience. This section explores the integration of KGs with LLMs and the benefits of NaPAs in education, focusing on increased understanding, personalization, and multimodal content manipulation and interaction.

\subsection{Knowledge Graphs in Education and Their Integration with LLMs}
\label{ssec:case-napa}
In educational applications, KGs can model the hierarchical \cite{ain2023automatic,chen2018knowedu} and associative nature of subject matter, allowing for a more detailed, structured approach to content delivery. This can enable accurate detection of the learner's current knowledge state and recommend the subject prerequisites that are missing from the learner's knowledge.
Educational KGs enable NaPAs to leverage background knowledge on a particular subject and understand the interconnections between various concepts within that subject. This understanding allows for adaptive and deeply personalized curriculum development, where instruction is continuously tailored according to the learner's progress and existing knowledge base.

Integrating KGs with LLMs offers significant advantages. While LLMs can process vast amounts of data to generate contextually relevant responses, they are often limited in their ability to consistently reason through complex problems or understand deeper semantic relationships within data with high contextual and factual accuracy. By incorporating KGs, the proposed hybrid system can reference structured, domain-specific knowledge as it interacts with learners, leading to more accurate, informed, and contextually relevant interactions. For instance, a student curious about a biology research could ask a question about cellular functions, prompting the agent to reference a KG structured with information about cell biology to provide a response that is both accurate and appropriate for the student’s grade level and prior understanding. 

Retrieval-Augmented Generation (RAG) \cite{lewis2020retrieval} further optimizes this integration by combining the generative abilities of LLMs \cite{gao2023retrieval} with the precision of KGs. In RAG models, relevant information is retrieved from a KG and is incorporated into the response generated by the LLM, producing answers that are not only linguistically coherent but also grounded in more accurate knowledge. This hybrid-optimized approach ensures that educational content remains reliable while still offering the flexibility and responsiveness needed to engage students in real-time. Such a setup could be used in any domain, where the system retrieves context-specific information to answer learners’ questions accurately, possibly correcting misconceptions and connecting various interrelated concepts seamlessly.

\subsection{Reasoning, Understanding, and Personalization in NAI Applications}
\label{ssec:case-nai-pur}
One of the key benefits of NAI in education is its capacity to enhance the reasoning abilities of PAs beyond the capabilities of current LLMs or KGs in isolation \cite{garcez2023neurosymbolic}. In the literature around PAs, we found that many PAs may provide responses based on limited data or pre-set algorithms, which means that they may lack the capabilities required to engage in complex reasoning about the topic, the learner, and the situation to provide deep personalization. NAI, however, can combine the interpretive abilities of neural systems with symbolic reasoning, allowing NaPAs to perform deeper personalization, tailoring all aspects (e.g., pedagogical approach, content being presented, modality of content, or pacing of the content presentation) of the educational experience to the individual needs of each learner. NAI allows NaPAs to interpret the nuances of student's emotional and knowledge state, recommend a learning path to fill the learner's knowledge gaps based on their learning needs (e.g., for example sourced from the personal KG).

As an example of the deep personalization possible with NaPAs, consider the unique advantage of NAI to offer recommendations for alternative actions to learners based on their current state (e.g., take a small break) and provide multimodal educational content. Educational content is often restricted by format; for instance, learners may generally only have access to written notes or video lectures, which may not suit all learning situations. NaPAs can overcome this limitation by dynamically generating and translating content across modalities, making learning materials more adaptable and inclusive. For instance, NaPAs could generate audio versions or podcasts of written material upon request, enabling learners to listen to lectures or notes. Alternatively, NaPAs are able to transcribe audio lectures into written notes. NaPAs could also create visual aids or diagrams to accompany textual information, providing alternative formats for understanding complex concepts through visual representations. Another major potential benefit of the multimodality of NaPAs and their abilities to convert information between modalities is in regards to accessibility. For example, NaPAs could convert visual graphs into narrated descriptions for the visually impaired or add captions to video lectures for hearing-impaired learners. This multimodal adaptability represents a significant leap forward in making education more inclusive, accessible, and aligned with diverse learning needs.

While the multimodal aspects of NaPAs are notable, their ability to communicate beyond the capabilities of LLMs alone offer additional potential benefits. The semantics from educational KGs~\cite{chen2018knowedu, qu2024survey} facilitate clear communication with the NAI system using parameters such as complexity level, preferred modality, contextual emphasis, etc., which can be leveraged to offer better personalized learning experiences. For instance, a learner could seamlessly navigate between a simple overview of a simplified textual summary, and an in-depth analysis for a comprehensive video lecture, and weave in custom examples, due to personalization features encoded in NaPAs. This semantic-driven customization ensures that the generated content aligns with the specific educational objectives and accessibility requirements of each learner. These improved multimodal and summarizing abilities also provide a solution within the realm of speech-language pathology, especially for individuals with speech and language impairments. For those individuals, where traditional text-based materials may pose challenges, this approach can transform these materials into accessible formats, facilitating both comprehension and engagement.

Another potential benefit of NAI is in regards to language: NAI could facilitate translation across languages, enabling a more inclusive learning environment for multilingual learners, learning a second language, or even for broader global content outreach. Imagine a NaPA within intelligent textbooks~\cite{jiang2023recent} that offers an automatic translation of a physics topic from English into Spanish, allowing learners who are not fluent in the primary language of instruction to follow along seamlessly. Further, NaPAs can also integrate with  intelligent textbook interfaces to offer connections between concepts across the entire body of knowledge connected within the educational KG, while also keeping in mind current learner objectives, allowing seamless understanding of the connections within the knowledge structure and identify any gaps in their comprehension for more targeted learning.

Another powerful feature of NaPAs is their ability to simulate nuanced, meaningful interactions, engaging learners in activities that based on their current state will challenge them and expose them to new learning objectives. By utilizing multimodal and dynamic interactions, the NaPAs will facilitate simulations of debates, critical questioning, and collaborative problem-solving, fostering deeper engagement and enhancing critical thinking in an innovative and transformative manner. For example, in a science course, a NaPA could dynamically adapt to a learner’s hypothesis about an experiment by analyzing their reasoning process in real-time, providing targeted feedback, and introducing unforeseen variables or counterarguments to deepen understanding. This personalized and responsive interaction replicates high-level cognitive tasks far beyond the capabilities of static, pre-programmed systems.

In summary, the key idea is that, while LLMs are capable of these various actions in isolation, they lack agency. The NaPA provides a mechanism by which an LLM, for example, can be leveraged to produce the various multimodal or multilingual adaptations on demand, based on its understanding of a learner -- either through an interpretation of an in-context student model or through the learner's personal KG. It then would be capable of grounding itself in literature, avoiding confabulation in this sensitive context, and further be capable of itself adapting to any domain where an educational or curriculum KG exists. These NaPAs could then be deployable on-demand to underserved populations, requiring only a device and internet connection for personalized learning.\footnote{We note that there is still a significant hurdle in achieving those requirements! However, initiatives such as \emph{One Laptop Per Child} \cite{site:olpc} or Project Connect Unicef \cite{site:unicef-connect}.}

\section{Preliminary Explorations}
\label{sec:results}
In our investigation into the application of NAI, we conducted a series of exploratory studies \cite{site:nai-explr} using various LLMs to generate educational content \cite{huang2024generating} and design curriculum related to KGs. The aim was to assess the capabilities of current models and evaluate their understanding of context. Our approach involved zero-shot content generation as a baseline, persona-based customization, and the integration of RAG techniques to enhance contextual relevance, evaluating the various components in the proposed system (Figure~\ref{fig:napa-ov1}). Additionally, we examined how this system could redesign curricula and learning paths, both with and without explicit incorporation of specific pedagogical strategies.

We initiated our exploration by prompting the LLMs to generate educational module content without providing any specific context or guidance—a process known as zero-shot generation \cite{liu2023pre, kojima2022large}. This approach allowed us to evaluate the LLMs’ inherent abilities to produce coherent and relevant educational material. Recognizing the importance of tailoring educational content to diverse learner profiles, we introduced persona contexts into the prompts. Personas were defined to represent various audience groups, such as college learners, designers, or business stakeholders. The system was tasked with generating module content tailored to these specific personas. This strategy enabled us to observe how the system adapted content to meet the unique needs and preferences of different audiences.

To enhance the contextual accuracy, and depth of the generated content, and maintain uniform coverage in the generation process, we employed and optimized RAG for our KG data \cite{lewis2020retrieval}. By integrating RAG, based on queries to a (manually curated) KG of curriculum, we observed significant improvement in the quality and relevance of the educational modules produced.

We then explored the system's ability to redesign existing curricula \cite{hu2024teaching} and refine learning paths without explicit prompts to incorporate pedagogical strategies. The system was provided with a generic curriculum and tasked with reorganizing it to better suit a specified persona. This exercise allowed us to evaluate the system's capacity to intuitively enhance curriculum structure based on learner profiles.

Building upon this previous exploration and with the intent of progressing towards a prototype of the proposed system, we introduced explicit prompts instructing the system to redesign curricula and learning paths while incorporating established pedagogical techniques. By guiding the system to integrate these methods, we aimed to assess its ability to produce curricula that not only align with learner needs but also adhere to evidence-based educational practices.

Our explorations revealed that the system demonstrated the ability to organize existing curriculum modules by introducing prerequisite modules and structuring learning paths that revisited broader topics very often. This approach facilitated connections between new and previously covered material, promoting knowledge reinforcement. Additionally, the system exhibited the ability to generate entirely new curricula and learning paths tailored to each persona, defining specific topics for each module and employing a repetition style that reinforced learning by revisiting key concepts.

Another notable characteristic of the system was its ability to adapt language to suit different personas, thereby maximizing content absorption by learners. These observations underscore the system's diverse methodologies in educational content generation and curriculum design. The integration of persona contexts and pedagogical strategies further enhanced the quality and relevance of the outputs, highlighting the importance of context-aware prompting in leveraging AI systems for educational applications.

Building on these advancements, NaPAs offer practical solutions to several challenges in education. They enable scalability by providing personalized instruction at scale, addressing resource constraints in traditional systems. NaPAs also serve as valuable assistants to educators by managing routine tasks like content creation and assessment, thereby freeing teachers to focus on deeper student engagement. Moreover, their multimodal capabilities enhance accessibility, ensuring inclusive education for learners with diverse needs, by having the capability to deliver content in various formats.

While the potential benefits of NAI in education are significant, several limitations and challenges must be addressed. Data privacy and security are critical concerns, as safeguarding sensitive student information is essential to maintaining trust. Bias and fairness present another key issue, requiring efforts to mitigate biases in AI systems to ensure equitable learning experiences. Teacher training and acceptance are equally important, as educators need to be trained and supported to effectively use AI tools. Finally, continuous improvement is vital, as ongoing evaluation and refinement are necessary to adapt to evolving educational and technological needs.

These preliminary findings from our exploration reveal several promising implications for the application of NAI systems in education. The current exploration highlights the remarkable potential for personalization, which can be further enhanced by integrating the system into a PA embodiment to strengthen the social dimensions of learning, thereby improving learner engagement and outcomes. Additionally, the proposed system is designed to become more adaptable as the learner progresses, with the inclusion of personal KGs \cite{balog2019personal} further refining its capacity for improved content delivery. By leveraging its demonstrated potential for tailored content delivery and the capabilities of contemporary models, this system holds promise for seamless delivery across diverse linguistic and multimodal formats. With the optimal integration of emerging NAI advancements, the proposed system is well-positioned to become a more adaptable learning solution in educational contexts.

\section{Conclusion}
\label{sec:conc}
NAI-augmented PAs will assist educators with routine (yet requiring significant effort or expertise) tasks while fostering deeper engagement and effective education for diverse learners, including those with disabilities and learning difficulties. These systems personalize and adapt learning by identifying knowledge gaps, tailoring instruction, and delivering multimodal, multilingual, inclusive content, and addressing challenges such as scalability, resource limitations, and accessibility. 

While we note that many of adaptive techniques for tailoring educational content for individual learners exist in isolation, NaPAs will understand when to use the various methods to produce a sum greater than the constituent components. That is, understanding when a learner requires a different approach, a modified example, a revision in their prerequisite knowledge, definitions in their native language, or a new modality, is a key advancement driven by the augmentation of PAs with NAI.

This is a transformative approach to educational enhancement, leveraging the combined strengths of neural and symbolic AI to develop intelligent, adaptable, and personalized pedagogical tools. By addressing contemporary challenges and revolutionizing the learning experience, NAI holds the potential to make education more engaging, inclusive, and impactful. As we continue to explore and refine these technologies, it is critical to consider ethical implications, ensure equitable access, and maintain a focus on the ultimate goal: improving educational outcomes for all learners.

\medskip

\noindent\emph{Acknowledgment.} Chris Davis Jaldi and Cogan Shimizu acknowledge support from the National Science Foundation under award \#2333532 ``Proto-OKN Theme 3: An Education Gateway for the Proto-OKN (EduGate).'' Noah Schroeder acknowledges support from the National Science Foundation and Institute for Education Sciences under Grant DRL-2229612. Any opinions, findings, and conclusions or recommendations expressed in this material are those of the author(s) and do not necessarily reflect the views of National Science Foundation or the U.S. Department of Education. Icons in Figure~\ref{fig:napa-ov1} are designed by Dewi Sari, Freepik, Flat Icons, and itim2101 from Flaticon. 

\bibliographystyle{abbrv}
\bibliography{refs}
\end{document}